\documentclass{article}
\usepackage{graphicx} 
\usepackage{amssymb} 

\usepackage{avant} 
\usepackage{mathptmx} 

\usepackage{microtype} 
\usepackage[utf8]{inputenc} 
\usepackage[T1]{fontenc} 

\usepackage{geometry} 

\usepackage{tcolorbox}

\usepackage[svgnames,x11names,table]{xcolor}
\usepackage{hyperref}
\definecolor{gmitblue}{RGB}{0,0,150}
\hypersetup{
    colorlinks = true,
    linkcolor=gmitblue,
    filecolor=gmitblue,
    urlcolor=gmitblue,
    citecolor = green!60!black
}

\usepackage{fontawesome}
\let\orighref\href
\renewcommand{\href}[2]{\orighref{#1}{#2\,{\footnotesize\faExternalLink}}}
\let\orighyperlink\hyperlink
\renewcommand{\hyperlink}[2]{\orighyperlink{#1}{#2\,{\footnotesize\faExternalLink}}}
\let\origurl\url
\renewcommand{\url}[1]{\origurl{#1}\,{\footnotesize\faExternalLink}}

\title{Good practices for evaluation of synthesized speech}
\author{Erica Cooper, Sébastien Le Maguer, Esther Klabbers, Junichi Yamagishi}
\date{October 2025}

\begin{document}

\maketitle

This document is provided as a guideline for reviewers of papers about speech synthesis.  We outline some best practices and common pitfalls for papers about speech synthesis, with a particular focus on evaluation.  We also recommend that reviewers check the guidelines for authors written in the paper kit\footnote{\url{https://www.overleaf.com/latex/templates/interspeech-paper-kit/kzcdqdmkqvbr}} and consider those as reviewing criteria as well.  This is intended to be a living document, and it will be updated as we receive comments and feedback from readers.  We note that this document is meant to provide guidance only, and that reviewers should ultimately use their own discretion when evaluating papers.

\begin{tcolorbox}[colback=red!5!white,colframe=red!75!black]
\bf \hyperlink{reviewing_checklist}{A reviewing checklist}, also serving as a summary of this document, appears on the final page.
\end{tcolorbox}
\section{Appropriate choice of evaluation metrics}

In any scientific paper, the evaluations chosen should address and verify the claims being made.  As for some examples in speech synthesis, if the target of a paper is speaker adaptation, then it is necessary to evaluate the effectiveness of the speaker adaptation by e.g. some measure of speaker similarity.  If the target is to improve the pronunciation of certain kinds of words or phonemes in certain contexts, then the evaluation material should contain those words or phonemes in those contexts, and so on.  It is important to consider what the use case is for the synthesizer being developed in the paper, and to consider whether the use case was sufficiently considered and incorporated into the evaluation -- for example, if the target listeners are language learners, then language learners should participate in the evaluation.

\subsection{Basic MOS tests for naturalness and the ``no use case'' use case}

It is extremely common for speech synthesis papers to aim for general-purpose speech synthesis without targeting a specific use case, and to show that ``naturalness'' was improved over some other previous state-of-the-art ``general-purpose'' system using a Mean Opinion Score (MOS) \cite{p800} test for naturalness or quality on individual sentences in isolation.  While this is a very popular genre of paper that holds interest for the community in terms of advancing the state of the art, there is nevertheless an abundance of evidence that this kind of MOS test has become saturated \cite{quest2018,LEMAGUER2024,PERROTIN2025101747} and care must be taken to ensure that evaluations were conducted in a way that enables meaningful distinctions to be made (e.g., see Section \ref{stats} on statistical analysis).  Furthermore, in order to make a meaningful scientific contribution to the community, \textbf{we recommend that papers conducting mainly this kind of use-case-free basic MOS evaluation should especially include some analysis about {\em why} their system outperformed the previous one.}

Conversely, we would like to remind reviewers that not every speech synthesis paper falls into this category and that a paper should not necessarily be penalized for {\em not} including a basic MOS test for naturalness -- it is not always sensible to conduct this kind of evaluation, especially when a paper is addressing a task or language for which no comparable prior work exists.  \textbf{If your review makes the critique that a MOS test should have been conducted, please also make a recommendation for which other comparison systems should be included in the MOS test and why.}  A single MOS value obtained in isolation is not meaningful and cannot be compared to MOS values in other papers due to many sources of bias \cite{zielinski2008some}.

We would also like to note that \textbf{the concept of ``naturalness'' is under-specified,} since  impressions of naturalness largely depend on context and on listeners' subjectivity.  Additionally, this term is often used to describe the evaluation when the question asked to listeners in fact uses different wording.   Authors should ensure that they have a shared understanding with listeners about what is actually being evaluated.  It is also important for authors to add a reference for the protocol of MOS test that they are using.  For instance, tests for naturalness should reference the Blizzard Challenge evaluation protocol \cite{blizzard2012} and tests asking listeners about quality should reference ITU-T Rec. P.800 \cite{p800}.  Authors should also note any deviations they made from these protocols.  Reviewers should check that existing protocols have been correctly implemented, modifications are justified, and that newly-proposed protocols have been adequately described.

\subsection{Alternatives to MOS}

While the MOS listening test paradigm may currently be one of the most popular ones, AB/ABX pairwise comparison tests, best-worst scaling tests (BWS) \cite{wells2024experimental}, MUSHRA-like tests\footnote{We note that a true MUSHRA test for TTS does not exist since there are no anchors.  Without anchors, the spread of the scale is unqualifiable and, much like MOS, these MUSHRA-like tests are relative as well.} \cite{mushra}, and the Audience Response System (ARS) \cite{taannander2024revisiting} are some examples of other valid and established testing paradigms.  Authors may also design their own tests that are specific to their use case or that measure the specific phenomenon under investigation -- this is valid and encouraged, as long as the authors justify and validate their proposed evaluation.

\section{Listening test evaluations in papers}

Since human listeners are typically the end users of speech technology, papers about speech synthesis should include an evaluation by human listeners.  If a listening test evaluation is not included in the paper, sufficient justification should be provided for why not.  Some examples of reasonable justifications for not conducting a listening test include the case of synthesis a very low-resource language where not enough native listeners can be recruited, or the case where the aspect of synthesized speech being investigated can be objectively measured in a well-established way.

It is important for papers to include some basic information about any listening tests that were conducted.  This is just as important as information about datasets, training procedures, model hyperparameters, etc.\ and should not be omitted. Based on~\cite{clark2007statistical,wester15c_interspeech,lemaguer25_interspeech}, the following is a list of basic information about listening tests that is necessary to report:

\begin{itemize}
    \item How many listeners participated?
\item How many audio samples per system?
\item How many ratings per sample?
\item What statistical test was performed?
\item What was the question that listeners were asked to answer?
\item If an Absolute Category Rating (ACR) scale (such as MOS) was used, what was the range of the scale and what were the increments?
\end{itemize}

The following are some additional good practices for listening tests:
\begin{itemize}
    \item Since the evaluation of the naturalness of speech is affected by differences in the volume of speech, it is necessary to be careful about the volume differences between the systems being compared.  (One popular tool for volume normalization of audio samples is the \texttt{sv56demo}\footnote{\url{https://github.com/openitu/STL/tree/dev/src/sv56}} implementation of ITU P.56 \cite{sv56}.)
    \item Naturalness evaluations may also depend on speaker characteristics \cite{cooper2021voices}, i.e. how much a subject likes a speaker.  If the speakers differ between systems, it is important to note that differences in naturalness ratings are not necessarily due to a difference in the performance of the synthesis model -- subjective preferences about the speaker are likely to have become a confounding factor.
    \item Listeners should be qualified to evaluate the hypothesis being tested in the paper.  Enough description of the listener population should be provided in the paper for reviewers to evaluate this point.
    \item Not all listeners are cooperative, especially in an online crowdsourced setting, so it may be necessary to screen them \cite{buchholz2011crowdsourcing}.
\end{itemize}

In addition to checking whether the listening test specifications were clearly stated and good practices followed, reviewers should also consider whether the listening test design is sensible -- e.g., is the number of listeners recruited enough \cite{wester15c_interspeech} for the chosen testing protocol and to have sufficient statistical power?  Are the chosen statistical tests appropriate (see Section \ref{stats})?  Are baseline comparison systems relevant and sensibly chosen to evaluate the claims of the paper, and are they open-sourced or is there otherwise enough information provided to reproduce them?

Authors should also include some information about their chosen test material, and reviewers should evaluate whether the material chosen is experimentally valid and sufficient for testing the authors' claims.  Were the test sentences or passages held-out data from a database? Were the speech materials designed for a specific purpose, such as semantically-unpredictable sentences (SUS) for evaluating intelligibility?

Last but not least, we would like to make note once more that \textbf{MOS values cannot be meainingfully compared across different papers or separate listening tests} because of all of the contextual biases that arise in these tests; authors who want to compare their synthesizer to other ones should include these other systems in their own listening test rather than citing the MOS values from those other papers, and reviewers are advised to check that comparisons are made properly.

\section{Use of automatic quality predictors}

Objective evaluations can provide useful information that can supplement the information obtained by listening tests.  As automatic MOS prediction has recently gained popularity as a research topic, with several pretrained and open-source prediction models being made available \cite{nisqatts,utmos}, automatic quality prediction including MOS prediction has recently been making more appearances in speech synthesis papers as an objective evalution method.  In the case that such predictors are used, \textbf{it is important for authors to clearly state which prediction model was used and to justify the reason why it was chosen from among the currently-available options.}  Papers using MOS predictors for evaluation should also include a justification for doing so.  

One recently-observed pitfall is to call such predicted values ``MOS'' -- while these models were indeed trained to predict MOS values, the predicted values are not anyone's ``opinion'' as such, and need to be clearly differentiated from the MOS values obtained in a listening test to avoid confusing readers.  \textbf{These predicted values should therefore be referred to as ``predicted MOS,'' ``automatic MOS,'' or similar.}

Several of the popular open-source speech quality predictors were trained in a supervised manner on labeled MOS data and their predictive abilities are therefore constrained to the domain of the training data.  \textbf{There is evidence that MOS predictors may have difficulty generalizing to domains that differ from their training data (e.g, speech synthesis in a different language),} and that the extent of the differences between domains may be difficult to guess \cite{vmc2023}.  \textbf{When using an automatic quality predictor, care should be taken to ensure that the one chosen was trained for a reasonably well-matched domain to the domain of the synthesized speech under evaluation.}

\section{Statistical significance reporting} \label{stats}

In line with the Interspeech scientific reporting checklist included in the author paper template, it is necessary to include ``some measure of statistical significance of the reported gains or confidence interval.''  The following is a (non-exhaustive) list of appropriate statistical tests for different types of listening test data:

\begin{itemize}
    \item MOS and MUSHRA: these values are ordinal and therefore {\em non-parametric} statistical tests are appropriate, such as the \textbf{Wilcoxon signed rank test} \cite{clark2007statistical} or the \textbf{Mann-Whitney U test} \cite{rosenberg17_interspeech}.
    \item Pairwise comparisons: a \textbf{z-test} or a \textbf{t-test} can be used depending on the sample size; the \textbf{binomial test} or the \textbf{Clopper–Pearson confidence interval} can also be used as described in \cite{yasuda2024automatic}.
    \item ARS: Mann-Whitney U test for number of clicks
\end{itemize}

As investigated in \cite{PERROTIN2025101747}, appropriately-chosen non-parametric pair-wise statistical analysis as described above is sufficient, but corrections such as Bonferroni correction can limit the power of these tests to discover significant differences.  It is also valid to first perform a single statistical test including all factors under investigation and then perform post-hoc multiple comparisons according to those factors, an approach which addresses this limitation.

Note that claims that two methods or systems are ``equivalent'' must be approached with caution when justified by a statistical test that shows no significant difference.  Reviewers should check that the listening test was designed and conducted in a rigorous enough manner for this kind of claim to be meaningful.  For example, pairwise comparisons can reveal more fine-grained differences than MOS tests \cite{goldstein1995}, and the selection of challenging test material (e.g., the text sequences listed in \cite{parallelneural}) can also reveal more differences between systems.  Test samples can also be chosen based on acoustic characteristics to maximally differentiate systems \cite{PERROTIN2025101747}.

\section{Contributors}

This document is an outcome of the Dagstuhl seminar 25032, ``Task and Situation-Aware Evaluation of Speech and Speech Synthesis.''  We thank Christina Tånnander, Jens Edlund, Petra Wagner, and Olivier Perrotin for their valuable input.

\bibliographystyle{IEEEtran}
\bibliography{main}

\newpage
\pagestyle{empty}

\hypertarget{reviewing_checklist}{}%
\begin{center}
    {\textbf {\fontsize{15}{60}\selectfont Checklist for Speech Synthesis Reviewers}}
\end{center}

\begin{itemize}
    \item[$\square$] The chosen evaluation protocol(s) should address and verify the claims being made.
    \item[$\square$] The synthesis use case addressed in the paper should be incorporated in the evaluation.
    \item[$\square$] Papers targeting ``no use case'' synthesis with basic MOS evaluations should especially take care to ensure that their evaluation protocol can reveal meaningful distinctions between systems, and should especially include some analysis about {\em why} their system outperforms baselines.
    \item[$\square$] Papers should {\em not} necessarily be penalized for not including a MOS test.  Reviewer recommendations to include a MOS test should also include recommendations for relevant comparison systems.
    \item[$\square$] ``Naturalness'' is under-specified and authors should ensure that they share an understanding with listeners about what to evaluate.
    \item[$\square$] Existing protocols should be referenced properly, and new protocols should be described in sufficient detail.  Alterations to existing protocols should be justified and described in sufficient detail.
    \item[$\square$] Descriptions of listening tests should include the following minimum necessary information: how many listeners participated, how many audio samples per system, how many ratings per sample, the type of statistical test performed, the question asked to the listeners, and the rating scale and its increments if one was used.
    \item[$\square$] Listeners should be qualified to evaluate the hypothesis, and the listener population should be sufficiently described.
    \item[$\square$] Enough listeners should be recruited for the listening test to have sufficient statistical power.
    \item[$\square$] An appropriate statistical test should be conducted.
    \item[$\square$] Baseline comparison systems should be sufficiently described for reproducibility, and sensibly-chosen to evaluate the claims of the paper.
    \item[$\square$] The test material chosen should be described and sensibly-chosen to evaluate the claims made by the paper.
    \item[$\square$] MOS values cannot be meaningfully compared across different papers or across separate listening tests.
    \item[$\square$] Use and choice of automatic quality predictors should be clearly stated, well-justified, and sufficiently described.
\end{itemize}

\end{document}